\documentclass[twoside]{article}
\usepackage{fleqn,espcrc2,amsmath,amssymb}

\usepackage{graphicx}
\title{Asymptotically free theories\ based on discrete subgroups
\thanks{Work supported in part by Schweizerischer
Nationalfonds} \thanks{Talk presented by F. Niedermayer} }

\author{Peter Hasenfratz\address[MCSD]{Institute of Theoretical Physics, 
       University of Bern, Sidlerstrasse 5, CH-3012 Bern, Switzerland}
	and Ferenc Niedermayer\addressmark[MCSD]\thanks{On leave from 
        the Institute of Theoretical Physics, 
        E\"otv\"os University, Budapest}}
       
\begin{document}

\begin{abstract}
We study the critical behavior of discrete spin models 
related to the 2d O(3) non-linear sigma model.
Precise numerical results suggest that models with sufficiently large
discrete subgroups are in the same universality class as the
original sigma model.
We observe that at least up to correlation lengths $\xi\approx 300$ 
the cut-off effects follow effectively an $\propto a$ behaviour both 
in the O(3) and in the dodecahedron model.

\end{abstract}

\maketitle

\section{Introduction}

Here we summarize some results of a recent paper \cite{unexp}
where discretized version of 2d O(3) sigma model and of 4d SU(N)
(N=2,3) gauge models were studied near their critical points.
Here we discuss only the 2d sigma model. For further results
and a more extensive list of references we also refer to that paper.

We consider models with discretized spin variables
taking values from the vertices of regular polyhedra.
The symmetry group is accordingly reduced from O(3) to the
corresponding discrete subgroup. 
At strong coupling (i.e. large temperature), the fluctuations are large,
hence the effects of restricting the spins to a discrete set 
are expected to be small.
However, at sufficiently small coupling the discrete system has to be
frozen. Decreasing $g$ from the strong coupling regime one expects
a second order phase transition at some value $g_c$.
We investigate the quantum field theory obtained in the limit
$g \searrow g_c$ and suggest that for sufficiently large subgroups
(icosahedron and dodecahedron) the resulting theory is the same 
as the one given by the original O(3) model in the $g\to 0$ limit.

The latter being asymptotically free (AF) this sounds surprising.
Our conjecture that the discrete model is AF should be
understood in the sense that the {\em physical running coupling} 
goes to zero as the corresponding momentum scale goes to infinity.
Here we study only the question whether the discrete model is
equivalent to the original O(3) model (which we can establish
only numerically, within our numerical precision, of course).
The title refers to the standard wisdom that the O(3) model is AF
-- a belief debated for long time by Patrascioiu and Seiler
(for references see \cite{PS}). Here we do not study this question.

The discrete subgroups were introduced \cite{Rebbi}
in the SU(2) Yang-Mills theory as an approximation thought to be valid
until the Wilson loops measured at the same {\em bare} coupling 
start to disagree. In contrast to this we compare the theories
with discrete and continuous groups at the same (large)
{\em correlation length}.
That a discrete symmetry of the action can be enhanced to the
continuous group at the criticality is known for the case 
of the 2d XY-model \cite{Jose}.

The investigations reported in ref.~\cite{unexp} and here,
were inspired by the works of Patrascioiu and Seiler on the
dodecahedron model \cite{PS} where they observed that it behaves
as the O(3) model. 
We investigated the conjecture that the discrete and continuous 
models are in the same universality class to a high precision,
down to O(0.1\%).

We consider here discretized spin models when the spin vectors point
to the vertices of a regular polyhedron embedded into the sphere
$S_2$. If not stated otherwise, we consider the standard nearest
neighbour (nn) action.
The spin models with small subgroups of O(3)
(tetrahedron, octahedron and cube, with 4,6 and 8 directions,
respectively) are {\em not} equivalent to the original O(3) model.
The tetrahedron is equivalent to a $q=4$ Potts model, while
the cube to 3 independent Ising models. 
Besides the standard O(3) action we also compared some of the results
to those for the O(3) fixed point action \cite{HN}.

We have chosen physical quantities which can be measured
with high precision: the finite size scaling function (FSSF) , the
renormalized zero momentum 4-point function $g_R$ and the same
coupling $g_R(z)$ in a finite physical volume. It is the latter
quantity which can be measured with the highest precision.

\section{The finite-size scaling function}

We consider an $L \times L$ periodic box and measure the 
second moment correlation length $\xi(L)$. At the same inverse
coupling $\beta$ we also measure the quantity $\xi(2L)$ 
on a $2L \times 2L$ lattice, and study the ratio $\xi(2L)/\xi(L)$
as a function of $\xi(L)/L$. 
The techique of finite size scaling (FSS) was used very effectively
by L\"uscher, Weisz and Wolff\cite{LWW} to obtain
the running coupling in the O(3) model.
Instead of the (exponential) correlation length defined in a strip
$L \times \infty$ used in \cite{LWW} we use the second moment
correlation length defined in a square box by
\begin{equation}
\label{xiL}
\xi(L)=\frac{1}{2\sin(\pi/L)}\sqrt{\frac{G_2(0)}{G_2(k_0)}-1}\,,
\end{equation}
where $G_2(k)$ is the 2-point spin correlation function in Fourier
space and $k_0=(2\pi/L,0)$.
The finite-size scaling function for this quantity has been measured
by Caracciolo et al.\cite{Car} in the same  model for a large set 
of $\xi(L)/L$ values.
We have chosen to measure around the value $\xi(L)/L\approx 0.4$.

\begin{figure}[htb]
 \begin{center}
  \includegraphics[width=7.5cm]{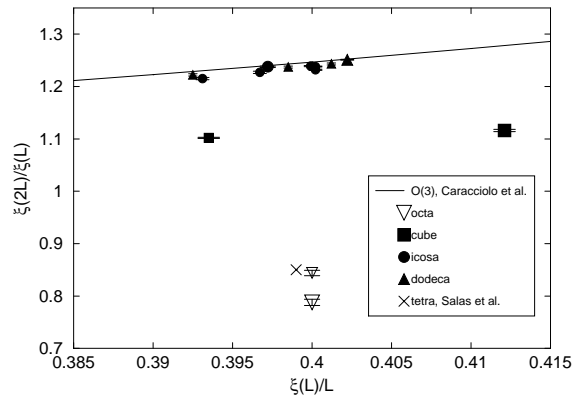}
  \caption{{}Part of the finite-size scaling curve (solid line) of 
   the O(3) non-linear $\sigma$-model~\cite{Car} is compared with
   the results of discrete spin models based on different discrete 
   subgroups.}
   \label{fig:fss3}
 \end{center}
\end{figure}

Fig.~\ref{fig:fss3} shows the FSSF for O(3) and different subgroups.
Points with symbols of the same shape belong to the same subgroup 
at different values of $L$, with the largest symbol corresponding 
to the largest $L$.
While the tetrahedron, octahedron and cube deviate strongly
from the O(3) curve and move away with increasing $L$,
the icosahedron and dodecahedron results are close to O(3),
and at largest $L$ lie on the curve within the small statistical
errors.

\section{The renormalized zero momentum 4-point coupling $g_{\rm R}$}

Define the quantity $g_{\rm R}(z)$ as
\begin{equation}
\label{gR}
g_{\rm R}(z)=\left( \frac{L}{\xi(L)}\right)^2
\left(1+\frac{2}{N}-\frac{\langle({\bf M}^2)^2\rangle}
{\langle{\bf M}^2\rangle^2} \right), 
\end{equation}
where $\xi(L)$ is the second moment correlation length, eq.~(\ref{xiL}),
$z=L/\xi(L)$, $N=3$ for O(3) and ${\bf M}$ is the magnetization,
\begin{equation}
M^a=\sum_x  S^a(x) \,.
\end{equation}
The coupling $g_{\rm R}=g_{\rm R}(\infty)$ is defined as 
the infinite volume limit $z\to\infty$, where first 
the continuum limit $\xi\to\infty$ is taken for fixed $z$.
For the O(3) $\sigma$-model $g_{\rm R}$ has been calculated
using the form-factor bootstrap method: $g_{\rm R}=6.770(17)$ in
agreement with the MC result $g_{\rm R}=6.77(2)$~\cite{Bal}.

\begin{figure}[htb]
 \begin{center}
  \includegraphics[width=7.5cm]{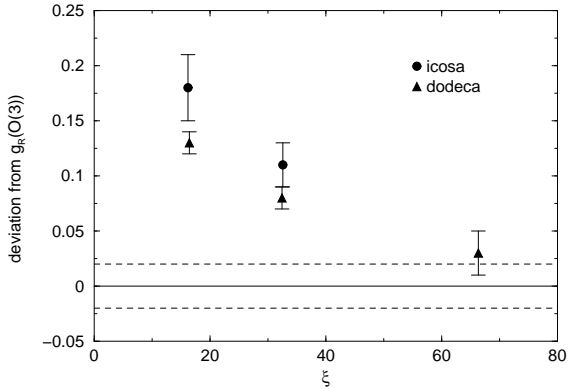}
  \caption{{}The deviation between the renormalized coupling of O(3)
  and that of the two largest subgroups as the function of the
  correlation length.}
  \label{fig:gr}
 \end{center}
\end{figure}

We have measured $g_{\rm R}(z)$ for the icosahedron and dodecahedron
models for $z\sim 6$ and extrapolated to infinite volume
using the finite size formula of ref.~\cite{Bal}. 
Figure~\ref{fig:gr} gives the deviation from the O(3) result as 
the function of $\xi$.
For comparison, the coupling from the cubic group is 
$g_{\rm R}^{{\rm cubic}}=1/3 g_{\rm R}^{{\rm Ising}}$ 
with $g_{\rm R}^{{\rm Ising}}=14.6975(1)$ from ref.\cite{Ising}, 
i.e. for this case the deviation from O(3) is 1.87(2).

As Figures~\ref{fig:fss3} and \ref{fig:gr} show, the two largest 
non-Abelian 
subgroups behave qualitatively differently from the small subgroups. 
The results suggest that the icosahedron and dodecahedron models are 
in the same universality class as the O(3) non-linear $\sigma$-model.

\section{The renormalized coupling in a finite volume}

The quantity $g_{\rm R}(z)$ can be measured with much better
accuracy for smaller z values as for large ones. 
In addition, one avoids the uncertainty by taking the infinite volume 
limit $z\to\infty$. 
We have chosen to measure in the vicinity of the arbitrary value
$z_0=2.32$.

Note that both $g_{\rm R}(z)$ and $z$ are measured in the same run.
Since they turn out to be strongly correlated the statistical
fluctuations can be strongly reduced by considering the quantity
$g_{\rm R}(z)-c(z-z_0)$ with an appropriate value of $c$.
Interpolating in $z\approx z_0$ gives $g_{\rm R}(z_0)$.
By measuring this quantity we could compare results for different 
models to a high precision.

Table~\ref{grdata} summarizes the results for the three actions
considered. 

\begin{table}[htbp]
  \begin{center}
\begin{tabular}{|rlll|}
\hline
$L$ & $\beta$ &   $z$     & $g_{\rm R}(z_0)$ \\
\hline
10  & 1.3637  & 2.3199(2) & 3.0105(1) \\
28  & 1.5785  & 2.3200(3) & 3.0765(2)  \\
56  & 1.697   & 2.3201(4) & 3.0979(2)  \\
112 & 1.8074  & 2.3230(4) &  3.1095(2) \\
224 & 1.9176  & 2.3178(5) &  3.1145(3) \\
448 & 2.028   & 2.3204(10)&  3.1175(6) \\
\hline
 10 & 0.8565 & 2.3202(4) &  3.1064(2) \\
 20 & 1.00   & 2.3197(5) &  3.1140(3) \\
 40 & 1.1347 & 2.3235(7) &  3.1163(4) \\
 80 & 1.2659 & 2.3199(9) &  3.1180(5) \\
\hline			 
 28 & 1.5636  & 2.3177(7) &  3.0683(5) \\
 56 & 1.670   & 2.3190(9) &  3.0923(7) \\
112 & 1.7626  & 2.3211(6) &  3.1051(5)\\
224 & 1.8465  & 2.3185(7) &  3.1106(5) \\
\hline
\end{tabular}
    \caption{{}Measurements of $z=L/\xi(L)$ 
     and $g_{\rm R}(z_0)$ for the O(3) nn action,
     the FP action and dodecahedron action, respectively.
     The last column was obtained by interpolating 
     $g_{\rm R}(z)-c(z-z_0)$ with $c=2$ and $z_0=2.32$.}
    \label{grdata}
  \end{center}
\end{table}

\begin{figure}[htb]
 \begin{center}
  \includegraphics[width=7.5cm]{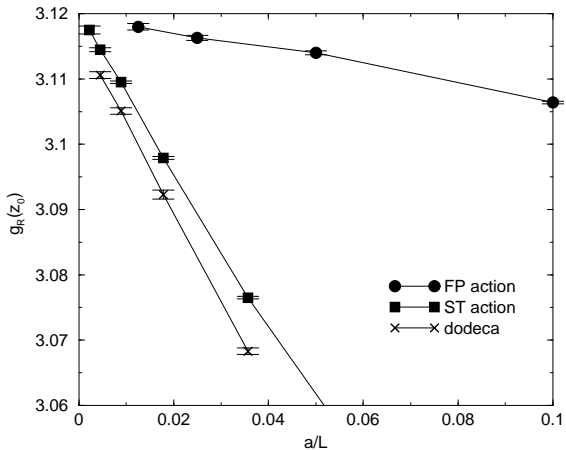}
  \caption{{}Cut-off dependence of $g_{\rm R}(z_0)$ (at $z_0=2.32$)
   for the standard, FP and dodecahedron actions.
   The data points are connected by straight lines to guide the eye.}
  \label{fig:gra}
 \end{center}
\end{figure}

Figure~\ref{fig:gra} shows $g_{\rm R}(z_0)$ as the function of $a/L$ for
the three actions considered.

The striking feature here is that even the standard O(3) action
shows a linear, O($a/L$), cut-off effect, while the perturbative 
approach predicts an  O($(a/L)^2$) behaviour (up to $\log(L/a)$
factors). 

Below we present for the O(3) nn action 
fits with different functional forms,
together with the corresponding value of $\chi^2/{\rm dof}$:
%\begin{alignat}{1}
%& 3.1201(2)-1.224(9)a/L\,,                     \nonumber \\
%& 3.1205(4)-1.28(5)a/L + 1.1(1.1) (a/L)^2\,,   \nonumber \\
%& 3.1207(6)-1.1(1)a/L - 0.04(4)a/L \log(L/a)\,,\nonumber \\
%& 3.1168(3)+103(5)(a/L)^2-40.4(1.4)(a/L)^2\log(L/a)\,, \nonumber
%\end{alignat}
\begin{alignat}{1}
& 3.1201(2)-1.224(9)\frac{a}{L}\,,                     \nonumber \\
& 3.1205(4)-1.28(5)\frac{a}{L} + 1.1(1.1)\left(\frac{a}{L}\right)^2\,, 
  \nonumber \\
& 3.1207(6)-1.1(1)\frac{a}{L} - 0.04(4)\frac{a}{L} \log\frac{L}{a}\,,\nonumber \\
 \nonumber
\end{alignat}
\begin{alignat}{1}
&
3.1168(3)+103(5)(\frac{a}{L})^2
 -40.4(1.4)\left(\frac{a}{L}\right)^2\log\frac{L}{a}\,,
 \nonumber
\end{alignat}

The corresponding  values of $\chi^2/{\rm dof}$ are 2.1, 2.2, 2.5 
and 4.1, respectively.
(The fits for the FP action and dodecahedron are given in
\cite{unexp}.)

One concludes that an $a/L$ term is needed to describe the cut-off
effects.  
(Note also the large coefficients in the $a^2$ fit!)

\section{On the cut-off effects}

For bosonic models in any order of perturbation theory 
the cut-off effects are given by O($a^2$) terms, up to 
logarithmic corrections \cite{Sym}.
Although not proven beyond perturbation theory, this form of
cut-off effects is generally assumed when one extrapolates to the
continuum limit.
Therefore an O($a$) is rather surprising.
Note that already in ~ref.~\cite{Bal} such behaviour
for $g_{\rm R}$ at $z\gtrsim 5$ was preferred. 
However, due to larger errors this form was not compelling.

Because the Ansatz for the cut-off corrections can significantly
modify the predictions in the continuum limit, and because
a non-standard behaviour of the lattice artifacts can have other
theoretical implications, it is important to investigate the related
questions further. In particular, it will be interesting to study
other non-perturbative quantities as  the LWW coupling \cite{LWW} 
in the O(3) sigma model.
Obviously, the same questions in d=4 gauge theories are even more relevant.

\eject

\end{document}